\documentclass[prb,superscriptaddress,showpacs,twocolumn,a4paper]{revtex4}

\usepackage{amssymb}
\usepackage{graphicx}
\usepackage{amsmath}
\usepackage{color}

\def\bk{\mathbf{k}}
\def\D{$\Delta$}

\def\x{\cos k_x}
\def\y{\cos k_y}
\def\z{\cos k_z}

\def\Hkin{\hat{H}_{\rm{kin}}}
\def\Hcf{\hat{H}_{\rm{cf}}}
\def\Hint{\hat{H}_{\rm{int}}}
\def\hepsk{\hat{\varepsilon}({\mathbf k})}
\def\hn{\hat{n}}
\def\dn{\delta n}
\def\Uc0{U_c^{\Delta=0}}
\def\Deff{\Delta_{\rm{eff}}}
\def\Umit{U_{\rm{MIT}}}

\def\e1{\epsilon_{1\mathbf{k}}}
\def\eps2{\epsilon_{2\mathbf{k}}}
\def\vk{V_{\mathbf{k}}}
\def\ek{e_{\mathbf{k}}}

\begin{document}

\title{Effect of Crystal-Field Splitting and Inter-Band Hybridization\\
on the Metal-Insulator Transitions of Strongly Correlated Systems}

\author{Alexander I. Poteryaev}
\affiliation{Centre de Physique Th{\'e}orique, UMR 7644, Ecole Polytechnique, CNRS, 91128 Palaiseau Cedex, France}
\author{Michel Ferrero}
\affiliation{Centre de Physique Th{\'e}orique, UMR 7644, Ecole Polytechnique, CNRS, 91128 Palaiseau Cedex, France}
\author{Antoine Georges}
\affiliation{Centre de Physique Th{\'e}orique, UMR 7644, Ecole Polytechnique, CNRS, 91128 Palaiseau Cedex, France}
\author{Olivier Parcollet}
\affiliation{Institut de Physique Th{\'e}orique, CEA, IPhT, CNRS, URA 2306, F-91191 Gif-sur-Yvette Cedex, France}

\date{\today}

\begin{abstract}
We investigate a quarter-filled two-band Hubbard model involving a crystal-field
splitting, which lifts the orbital degeneracy as well as an inter-orbital 
hopping (inter-band hybridization).
Both terms are relevant to the realistic description of correlated
materials such as transition-metal oxides.
The nature of the Mott metal-insulator transition is clarified and
is found to depend on the magnitude of the crystal-field splitting. At large
values of the splitting, a transition from a two-band to a one-band metal 
is first found as the on-site repulsion is increased and is followed by a Mott transition 
for the remaining band, which follows the single-band (Brinkman-Rice) scenario well 
documented previously within dynamical mean-field theory.
At small values of the crystal-field splitting, a direct transition from a 
two-band metal to a Mott insulator with partial orbital polarization is found,
which takes place simultaneously for both orbitals. 
This transition is characterized by a vanishing of the quasiparticle weight
for the majority orbital but has a first-order character for the minority orbital.
It is pointed out that finite-temperature effects may easily turn the metallic regime into a
bad metal close to the orbital polarization transition in the metallic phase.
\end{abstract}

\maketitle

\section{Introduction}
\label{sec:intro}

The Mott metal-insulator transition~\cite{Mott_book,Imada1998}
plays a key role in the physics of strongly correlated electron materials.
Over the last fifteen years, our theoretical understanding of this
phenomenon improved considerably, due to the development of the dynamical mean-field
theory (DMFT) (Refs.~\onlinecite{dmft_review} and \onlinecite{Kotliar2006}).
A number of model studies were performed in order to clarify the nature of the
transition in both a single-orbital and multi-orbital 
context~\cite{Pruschke2005,PhysRevB.66.115107,PhysRevB.55.R4855,Werner2007,Koch1999,Florens2002,Ono2003,Han1998,Laloux1994}.

In the context of real materials, however, several important features must be considered,
which are not always taken into account in model studies. This includes in particular
two key aspects: {\it (i)} the breaking of orbital degeneracy by the crystalline
environment and {\it (ii)} the existence of hopping terms coupling different orbitals
on different sites of the crystal (inter-orbital hopping or hybridization).
We note at this stage that the breaking of orbital degeneracy can correspond to a rather
large energy scale (of order 1-2 eV) when one has in mind the crystal-field
splitting between $t_{2g}$ and $e_g$ levels in a transition-metal oxide, but it can also
correspond to a smaller energy scale (a small fraction of an electron volt) when considering,
e.g., the trigonal splitting of the $t_{2g}$ levels induced by a distortion of the cubic symmetry.
In the former case, an effective model with fewer orbitals can often be considered, but in the
latter case, all orbital components may still be relevant, albeit with
different occupancies, and one has to use a model involving several orbitals with slightly different
atomic level positions. In the present paper, we shall designate the lifting of orbital degeneracy
by the generic term of ``crystal-field splitting,'' but it is mostly the case where this is
a small energy scale (e.g., trigonal splitting of the $t_{2g}$ multiplet) that we have in mind
for applications.

Indeed, the physical effects arising from the competition of crystal-field splitting
and strong correlations have attracted a lot of attention recently, in particular
in LDA+DMFT electronic structure studies of many different compounds.
We now quote just a few examples. Pavarini {\it et al.}~\cite{Pavarini2004}  pointed out
that the lifting of cubic symmetry by the GdFeO$_3$-type distortion plays a key role in
determining the metallic or insulating characters of $d^1$ transition-metal perovskites such as
(Sr/Ca)VO$_3$ (small distortion, metals) and (La/Y)TiO$_3$ (larger distortion, insulators).
Indeed, for a given value of the on-site Coulomb repulsion $U$, the lifting of the
orbital degeneracy makes the insulating state more easily accessible~\cite{Koch1999,Florens2002}. 
Furthermore,
correlation effects considerably enhance the effective crystal-field splitting, hence
favoring orbital polarization (as also emphasized in Ref.~\onlinecite{Mochizuki2003} for these compounds).
This correlation-induced enhancement of the effective crystal-field splitting and
this increased orbital polarization
have also been shown~\cite{Laad2003,Poteryaev2007} to play a key role in
the metal-insulator transition of V$_2$O$_3$, with the $e_g^\pi$ component of the
$t_{2g}$ level much more occupied than the $a_{1g}$ component in the insulating phase
(see also the previous LDA+DMFT studies of V$_2$O$_3$ in Refs.~\onlinecite{Keller2004}
and \onlinecite{Laad2006}).

Such effects were discussed, at a model level, in the pioneering paper of
Manini {\it et al.}~\cite{PhysRevB.66.115107}, motivated by the physics of
fullerene compounds. In this work, a model consisting of two orbitals occupied
by one electron (quarter-filling) was considered, and the combined effect of a crystal-field
and of on-site repulsion was studied in the framework of DMFT. This work identified several
phases, most notably a two-band metallic phase (with partial orbital polarization)
and a one-band metallic phase (with full orbital polarization), as well as a fully orbitally-polarized
Mott insulating phase.

However, some questions of great importance
were left unanswered by this early study. To quote just a few of these issues:
{\it (i)} What is the nature of the metal-insulator transition in the different ranges of crystal field?
{\it (ii)} How exactly does the crossover between a two-orbital Mott transition to 
a one-orbital Mott transition takes place?
{\it (iii)} What is the effect of an inter-orbital hopping, always present in real materials, and in particular
does it wipe out the two-band metal to one-band metal transition within the metallic phase?
and, finally, {\it (iv)}, is it possible to obtain within DMFT the insulating phase with partial
orbital polarization, which is expected from general strong-coupling arguments?
(As we shall see, the answer is affirmative
and this phase was overlooked in the DMFT study of Ref.~\onlinecite{PhysRevB.66.115107}).

All these questions are directly relevant to the understanding of real materials 
(e.g., V$_2$O$_3$ and Sr$_2$RuO$_4$)
and to a better qualitative interpretation of the results of LDA+DMFT studies.
The aim of the present article is to provide a detailed answer to these questions.
This is made possible, in particular, by the recent development of numerical techniques for
solving efficiently the DMFT equations, in particular 
continuous-time Monte Carlo algorithms~\cite{Rubtsov2004,Rubtsov2005,Werner2006,Werner2006a}.

Let us point out that another related model study recently appeared, namely that
of the two-orbital model at {\it half-filling} 
(i.e., two electrons in total)~\cite{Werner2007}.
In this case, the physical issues are quite different since one
evolves from a two-orbital Mott insulator in the absence of crystal field to a band insulator
at large crystal field (not a one-orbital Mott insulator as in our quarter-filled case). Also,
this study did not consider the effect of an inter-orbital hopping. In this respect, our work
and Ref.~\onlinecite{Werner2007} can be considered quite complementary to one another.

Finally, we emphasize that the interplay between crystal-field splitting and strong
correlations is made even more complex in the presence of Hund's coupling and exchange
terms. In a study of BaVS$_3$
it was pointed out that when Hund's rule wins over crystal-field effects, one can observe
a {\it compensation} between orbital populations rather than an enhanced orbital
polarization~\cite{Lechermann2005,Lechermann2005a,Lechermann2007}.
The competition between Hund's coupling and crystal-field is also
relevant to the physics of cobaltites~\cite{Marianetti2004,Ishida2005,Perroni2007,Marianetti2007}, 
ruthenates,~\cite{Liebsch2000,Anisimov2002,Liebsch2007,Dai2006} and monoxides
under pressure~\cite{Kunes2008}. In the present work, however,
we focus on the interplay of crystal field and strong correlations and on the
nature of the Mott transition, in the simplest possible context and consider only the
effect of an on-site repulsion.

This paper is organized as follows:
In Sec.~\ref{sec:model}, we introduce the model and some notations.
In Sec.~\ref{sec:phase_diagram} we present the phase diagram and discuss
qualitatively each phase.
In Sec.~\ref{sec:lowDelta}, we discuss in details the insulating phases, using both an
analytical strong-coupling method and complete numerical solution of the DMFT equations.
In Sec.~\ref{sec:highDelta}, we clarify the nature of the various phase
transitions: from a two-band to a one-band metal and from a metal to a Mott insulator,
in the different crystal-field regimes. Finally, in Sec.~\ref{sec:finiteV}, we consider
the effects of a finite inter-orbital hopping and also we discuss
some finite temperature effects in regimes where the two orbitals have very different
quasiparticle coherence scales.

\section{Model}
\label{sec:model}

We consider a minimal two-band Hubbard model with crystal-field splitting and inter-orbital
hybridization, given by  the Hamiltonian;
\begin{equation}
  \label{eq:H_general}
  \hat{H} = \Hkin + \Hcf + \Hint,
\end{equation}
with;
\begin{subequations}
\begin{eqnarray}
&&\Hkin = \sum_{\mathbf{k}}\sum_{\sigma mm'}
\hepsk_{mm'} d^\dagger_{\bk\sigma m} d_{\bk \sigma m'},
\label{eq:Hkin}\\
&&\Hcf = \frac{\Delta}{2} \sum_{i\sigma} (\hat{n}_{i\sigma 1} -\hat{n}_{i\sigma 2}),
\label{eq:Hcf} \\
&&\Hint = \frac{U}{2} \sum_i \sum_{m\sigma\neq m'\sigma'} \hn_{i\sigma m} \hn_{i\sigma' m'}.
\label{eq:Hint}
\end{eqnarray}
\end{subequations}
In these expressions, $i$ is a lattice-site index, $\bk$ is the momentum in reciprocal space,
$m=1,2$ is an orbital index, and $\sigma=\uparrow,\downarrow$ is a spin index.
The sum in the interaction term runs over all orbital and spin indices except
the case when $m=m'$ and $\sigma=\sigma'$ and therefore all intraorbital and 
inter orbital Coulomb interactions are included.
$\Delta$ is the crystal-field splitting
between the two orbitals ($\Delta>0$ favors the population of the second orbital, $m=2$),
and $U$ is the density-density Coulomb interaction between the two orbitals.

In this article, we focus on quarter-filling (i.e., one electron in two orbitals, per lattice site),
which is achieved by tuning appropriately the chemical potential $\mu$.
We consider only the density-density form of the interaction term, and we do not include the Hund's exchange,
spin-flip, or pair-hopping terms.
The motivation for neglecting these terms is to keep the Hamiltonian as simple as possible. Note however that,
with one electron per site, the effect of these terms is expected to be small and acts basically as
a renormalization of the on-site $U$ (Refs.~\onlinecite{Ono2003} and \onlinecite{Han1998}).

The kinetic term $\Hkin$ is a two-band tight-binding
Hamiltonian on the three-dimensional cubic lattice
(we will also use its Bethe lattice counterpart),
which can be written (in $\bk$ space) as
\begin{equation}
\hepsk = \biggl[
                  \begin{array}{ll}
                     e(\bk) &   V(\bk)  \\
                     V(\bk)  &  e(\bk)
                  \end{array}
                \biggr],
\end{equation}
where diagonal elements correspond to the simple cubic lattice,
and the off-diagonal ones have $x^2-y^2$ symmetry;
\begin{subequations}
  \label{eq:H_kin}
  \begin{eqnarray}
    \label{eq:ek}
    e(\bk) & = &    2 t ( \x + \y + \z),             \\
    \label{eq:vk}
    V(\bk) & = &    2 \sqrt{3} V ( \x - \y ) \z.
  \end{eqnarray}
\end{subequations}
This corresponds to a hopping between identical orbitals on nearest-neighbor
sites, equal to $-t$.
The inter-orbital hopping connects orbitals $m=1$ and $m=2$ on next-nearest-neighbor sites and
is equal in magnitude to $\sqrt{3} V / 2$.
It has a positive sign for the $[ \pm 1, 0, \pm 1 ]$ neighbors and negative
for the $[ 0, \pm 1, \pm 1 ]$ ones.
This symmetry choice insures that, for all values of $V$, the on-site ($\bk$ integrated) kinetic Hamiltonian is
{\it diagonal in orbital space}. This is also the case of all local ($\bk$ integrated) quantities
in the interacting model, as can be checked by expanding the Green's function in power of $V$.
Hence, our model is such that the choice of local orbital basis set is adapted to the
local crystal symmetry. Physically, the model [Eq.~(\ref{eq:H_kin})] is a reasonable description, for
example, of an $e_g$ doublet split by the breaking of the cubic symmetry.

For zero hybridization, $V$=0, the density of states (DOS) is reduced to
the DOS of the cubic lattice for both orbitals shifted by $\pm \Delta/2$.
We set the energy unit by $t=1/6$, or equivalently $D=1$, where $D$ is the half-bandwidth.

We solve this model in the DMFT framework~\cite{dmft_review}. Since our main aim is to elucidate
the nature of the metal-insulator transitions in this model, we focus in this
article on the paramagnetic phases.
The self-consistent impurity problem is solved with two numerical techniques: {\it (i)}
Exact diagonalization (ED) as described in (Refs.~\onlinecite{dmft_review} and 
\onlinecite{PhysRevLett.72.1545}),
with a ``star-geometry'' for the bath hybridization function
using five bath states per orbital degree of freedom;
{\it (ii)}
the recently introduced continuous time quantum Monte Carlo algorithm (CT-QMC)
using an expansion in the impurity model hybridization function~\cite{Werner2006,Werner2006a}.
CT-QMC is more precise than ED and is necessary to establish the existence of the partially
polarized insulator phase (see Sec.~\ref{sec:lowDelta}), as we shall discuss further below.

\section{Results in the absence of inter-orbital hybridization}

\subsection{Zero-temperature phase diagram}
\label{sec:phase_diagram}

\begin{figure}[t]
   \centering
   \includegraphics[clip=true, angle=270, width=.47\textwidth]{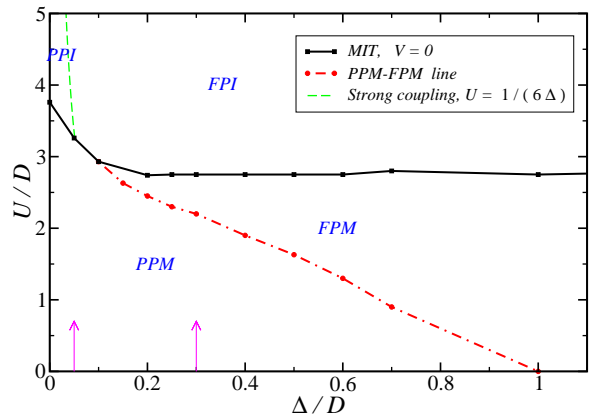}
   \caption{(Color online) Zero-temperature phase diagram (paramagnetic phases),
            on the cubic lattice without
            hybridization ($V=0$), and for one electron per site.
            The solid (black) line separates metallic and insulating regions.
            The dot-dashed (red) line separates the partially polarized metal (PPM)
            and the fully polarized metal (FPM) (for details see text).
            The dashed (green) line is the result of the strong-coupling mean-field analysis
            (see Sec.~\ref{sec:strong} and Eq.~\ref{eq:polarization_exact}): It
            separates the partially-polarized insulator (PPI) from the fully-polarized insulator (FPI).
            Arrows indicate the set of parameters used in
            Figures~\ref{fig:PPM-FPM-FPI} and~\ref{fig:PPM-PPI-Transition}.
            The ED solver was used.}
   \label{fig:phase_diagram}
\end{figure}
The DMFT phase diagram of model [Eq.~(\ref{eq:H_general})] at quarter-filling and without
inter-orbital hybridization ($V=0$) is presented on
Fig.~\ref{fig:phase_diagram}.
The effect of a non-zero $V$ will be considered in Sec.~\ref{sec:finiteV}.
The general shape of this phase diagram can be easily anticipated by considering
the various limiting cases~\cite{PhysRevB.66.115107}:

{\it (i)}
For $\Delta=0$, one has a well documented two-band degenerate model.
The model undergoes a correlation-driven Mott transition at a critical
$U_{c}^{\Delta=0}\simeq 3.76$
which is close to the results obtained by other authors for the Bethe lattice (semi-circular DOS with
identical half-width $D=1$)~\cite{Pruschke2005,PhysRevB.66.115107,PhysRevB.55.R4855}.

{\it (ii)}
For very large $\Delta\gg D$, the minority orbital (orbital $m=1$) is pushed to very high energy and
becomes completely empty, so that it can be ignored altogether.
The quarter-filled two-band model thus reduces to a single-band model {\it at half-filling}.
This situation has been thoroughly studied within DMFT and yields a correlation-induced
Mott transition at $U_{c}^{\Delta\rightarrow \infty} \simeq 2.75$.
(see, e.g., Ref.~\onlinecite{dmft_review} and references therein).
The metal-insulator transition line (plain/black line on Fig.~\ref{fig:phase_diagram})
interpolates between the limiting critical couplings corresponding to
$\Delta=0$ and $\Delta=\infty$. The system is insulating above this
line and is metallic below.

{\it (iii)} 
The non-interacting model ($U=0$) obviously has a transition between a
two-band metal for $\Delta<D$ and a one-band metal for $\Delta>D$. For $\Delta=D$, the minority band
crosses the Fermi level and becomes empty. This effective-band transition separating a
two-band situation at low energy from a non-degenerate band can actually be followed through
the phase diagram (dashed-dotted/red and dashed/green lines on Fig.~\ref{fig:phase_diagram}),
as we now discuss.

We note that we have not attempted to precisely determine
whether the orbital-polarization lines
cross the metal-insulator transition (MIT) line at a single point, 
or whether the orbital polarization line in the insulating
phase and in the metallic phase hit the MIT boundary at slightly different locations.

In the absence of hybridization ($V=0$), we can use the orbital polarization as a faithful indicator
of the transition between the two-band and a one-band regime. This quantity is defined as
\begin{equation}
  \dn = \frac{\langle \hn_{>}\rangle-\langle \hn_<\rangle}
             {\langle \hn_{>}\rangle+\langle \hn_<\rangle},
\end{equation}
in which $>$ and $<$ stand for the majority and minority orbitals, respectively,
($\langle \hn_{>}\rangle > \langle \hn_<\rangle$). At quarter filling and
for $\Delta>0$, this reduces simply to $\dn = \langle \hn_2 -\hn_1\rangle$.

As the crystal-field splitting is increased, one reaches a critical value at which
the orbital polarization reaches $\dn=1$, indicating a completely empty minority
orbital. The line along which this happens in the $(\Delta,U)$ plane, is indicated by
the dashed-dotted (red) line in the metallic phase and by the dashed (green) line in the
insulating phase. Hence, four different phases are apparent on the phase diagram
of Fig.~\ref{fig:phase_diagram}: a partially polarized (two-band) metal (PPM), a
fully polarized (one-band) metal (FPM), a partially polarized Mott insulator (PPI), and a
fully polarized Mott insulator (FPI).

As already pointed out by Manini {\it et al.}~\cite{PhysRevB.66.115107},
and as clear from Fig.~\ref{fig:phase_diagram},
the value of the crystal-field, at which the transition
from the PPM to the FPM takes place, is strongly
reduced by interactions. While it is set by the half-bandwidth at $U=0$, it is renormalized down
by the quasiparticle weight $Z_>$ in the presence of interactions. 
Hence, a crystal-field splitting considerably
smaller than the half-bandwidth can be
sufficient to induce a two-band to one-band metal transition.

It is important to realize, however, that the value of $\Delta$ needed to fully polarize
the system vanishes only in the limit $U=\infty$.
In other words, the orbitally-degenerate Mott insulator at $\Delta=0$ has a {\it finite orbital
polarizability}, even within the DMFT approach. Hence the PPI phase at large $U$ and
small values of $\Delta$ exists. This point was incorrectly appreciated by
Manini {\it et al.}~\cite{PhysRevB.66.115107}, largely for numerical reasons.
Indeed, the ED algorithm is inappropriate to correctly capture the PPI phase.
In the present article, we establish (Sec.~\ref{sec:lowDelta}) the existence of the partially
polarized insulating phase within DMFT using both an analytical proof at strong-coupling limit and
a complete numerical solution of the DMFT equations based on the new CT-QMC algorithm.

Let us point out that in this zero-temperature phase diagram, all the transitions are
second order, except for the transition from the PPM to the PPI -- which is second order
for the majority orbital and first order for the minority orbital, as will be explained
below in Sec.~\ref{sec:MITsmallDelta}. At finite temperatures $T > 0$, the MIT becomes
first order throughout the phase diagram, as in canonical DMFT solutions, whereas
the other transitions remain second order.

In the two following subsections, we describe in more details the nature of these
different phases and we investigate the phase transitions between them.

\subsection{Existence of the partially polarized insulator}
\label{sec:lowDelta}

\subsubsection{Strong-coupling analysis: Kugel-Khomskii model}
\label{sec:strong}

At strong coupling $U\gg D$ (or $U \gg t$), in the Mott insulating phases, an effective low-energy model
can be derived, following Kugel and Khomskii~\cite{Kugel1982} (see also Ref.~\onlinecite{PhysRevB.52.10114}).
The low-energy Hilbert space contains only the four states $|i,m,\sigma\rangle$ with one electron
on each site ($m=1,2\,$;\,$\sigma=\uparrow,\downarrow$).
The effective Hamiltonian acting on these states reads,
\begin{multline}
  \hat{H}_{\rm{eff}} = - \Delta \sum_i \hat{T}_i^z +         \\
                    + \sum_{\langle ij \rangle} \Big\{ J_s ( \vec{S}_i \vec{S}_j ) +
                                       J_o ( \vec{T}_i \vec{T}_j ) +
                                       J_m ( \vec{S}_i \vec{S}_j ) ( \vec{T}_i \vec{T}_j ) \Big\}.
\label{eq:H_eff}
\end{multline}
In this expression, $\langle ij\rangle$ denotes the bonds between nearest-neighbor sites, and the
spin and pseudo-spin (i.e., orbital isospin) operators are given by:
\begin{subequations}
  \begin{eqnarray}
    &&\vec{S}_i=\frac{1}{2} \sum_m d^\dagger_{i\sigma m}\vec{\tau}_{\sigma\sigma'} d_{i\sigma' m}, \\
    &&\vec{T}_i=\frac{1}{2} \sum_\sigma d^\dagger_{i\sigma m}\vec{\tau}_{m m'} d_{i\sigma m'},
  \end{eqnarray}
\end{subequations}
in which $\vec{\tau}$ are the Pauli matrices. In particular, the $z$ component of these operators
(with eigenvalues $\pm 1/2$) is given by:
\begin{subequations}
  \begin{eqnarray}
    &&\hat{S}^z_i = \frac{1}{2} ( \hat{n}_{i\uparrow 2} - \hat{n}_{i\downarrow 2} +
                                  \hat{n}_{i\uparrow 1} - \hat{n}_{i\downarrow 1} ),
  \label{eq:Sz}\\
    &&\hat{T}^z_i = \frac{1}{2} ( \hat{n}_{i\uparrow 2} + \hat{n}_{i\downarrow 2} -
                                  \hat{n}_{i\uparrow 1} - \hat{n}_{i\downarrow 1} ).
  \label{eq:Tz}
  \end{eqnarray}
\end{subequations}
The (superexchange) couplings $J_s$, $J_o$, and $J_m$ are given
by~\cite{PhysRevB.52.10114}
\begin{equation}
  J_s = J_o = \frac{J_m}{4} = \frac{2t^2}{U} \equiv J.
  \label{eq:J}
\end{equation}
The particular symmetry between these couplings is due to the choice of a
density-density interaction and to the neglect of the Hund's exchange.

At strong coupling, in the insulating phase, the DMFT solution of the
original model [Eq.~(\ref{eq:H_general})] reduces to a static mean-field solution
of Eq.~(\ref{eq:H_eff}). Focusing on the non-magnetic phase
($\langle S^z_i\rangle =0$), the orbital polarization
$\dn = 2\langle T^z\rangle$ is given by the self-consistent equation, at
finite temperature $T=1/\beta$:
\begin{equation}
  \delta n = \tanh \Big( \frac{\beta}{2} ( \Delta - \Delta_c \delta n) \Big),
  \label{eq:polarization_exact}
\end{equation}
where
\begin{equation}
  \Delta_c = \frac{zJ}{2}=z \frac{t^2}{U},
\end{equation}
is a critical value of the crystal-field splitting and $z$ is the coordination
number of the lattice (number of nearest neighbors).
For the simple cubic lattice, with $z=6$ and half-bandwidth $D=zt$, this yields,
$\Delta_c^{CL} = D^2/(6U)$, while for the large-connectivity Bethe lattice with
nearest-neighbor hopping $t=D/(2\sqrt{z})$, one has: $\Delta_c^{BL} = D^2/(4U)$.

At zero temperature ($\beta=\infty$), the solution of Eq.~(\ref{eq:polarization_exact}) reads,
\begin{equation}
  \label{eq:polarization_T=0}
  \dn =
    \begin{cases}
       \;\;  \Delta / \Delta_c, & \quad \Delta < \Delta_c  \\
       \;\;     1,              & \quad \Delta \geqslant \Delta_c
    \end{cases}.
\end{equation}
Hence, this shows that the orbitally degenerate insulator has a finite orbital susceptibility
at $T=0$, $\chi_{\rm{orb}}=1/\Delta_c$, and that a finite crystal-field $\Delta=\Delta_c$ must be
applied to fully polarize the insulating phase. The strong-coupling expression
$\Delta=\Delta_c=D^2/(6U)$ for the cubic lattice corresponds to the dashed(green) line displayed
on Fig.~\ref{fig:phase_diagram}, separating the PPI from the FPI phases at $T=0$.

At finite temperature, a good approximation to the solution of Eq.~(\ref{eq:polarization_exact})
turns out to be;
\begin{equation}
  \delta n \simeq \frac{\Delta}{\Delta_c} \frac{1}{ 1 + \frac{2}{\Delta_c\beta}}.
  \label{eq:polarization_apprx}
\end{equation}

Finally, we would like to emphasize that, when thinking of DMFT as an exact method
in the limit of large lattice coordination $z\rightarrow\infty$, it is quite clear that a non-zero
value of the critical $\Delta_c$ (and hence a finite extent of the PPI phase) is to
be expected. Indeed, the orbital exchange coupling [Eq.~(\ref{eq:J})] scales
as $1/z$ (since $t\propto 1/\sqrt{z}$), hence, the critical $\Delta_c$ is of
the order of the exchange field between a site and all its neighbors, i.e., of order $zJ$, which
remains $O(1)$ as $z\rightarrow\infty$. The {\it uniform} orbital susceptibility of the
orbitally degenerate Mott insulator is indeed finite at $T=0$ (this should not be confused with the
fact that the {\it local} susceptibility would scale as $1/z$ and hence vanish in the
large-$z$ limit). 

\subsubsection{Numerical solution: Importance of global moves in
the quantum Monte Carlo algorithm}
\label{sec:numerical}

\begin{figure}[t]
   \centering
   \includegraphics[clip=true, angle=270, width=.475\textwidth]{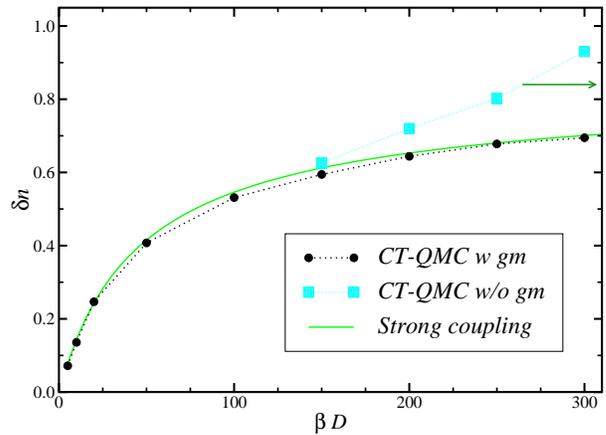}
   \caption{(Color online) Temperature dependence of the orbital polarization,
            $\delta n$, for the cubic lattice in the PPI phase.
            Black dots are the CT-QMC data with the use of global moves ({\it gm})
            in the spin and orbital space (see text for details).
            (Cyan) squares are the CT-QMC data without global moves.
            The solid (green) line is the strong coupling result given by
            Eq.~\ref{eq:polarization_exact}. The arrow shows the zero-temperature
            value of the polarization in the strong coupling limit, $\delta n$ = 0.84.
            These results are obtained for $U$=4, $\Delta$=0.035 and using the CT-QMC solver. }
   \label{fig:dn_vs_beta}
\end{figure}

The analytical estimate at finite temperature
[Eq.~(\ref{eq:polarization_apprx})] provides a very useful benchmark when
solving numerically the DMFT equations for the original model in the
strong-coupling regime. Indeed, it is actually non-trivial, from the
numerical point of view, to successfully stabilize the partially polarized
insulating phase. To achieve this, we have used the CT-QMC method, and it proved
necessary to implement global Monte Carlo moves,
in addition to the Monte Carlo moves proposed in Ref.~\onlinecite{Werner2007}.
In CT-QMC, a configuration is given by a collection of fermionic operators
$c_{\alpha_1} (\tau_1) \ldots c_{\alpha_N} (\tau_N)$ at different
imaginary times $\tau_i$ and the $\alpha_i$ are the fermionic species of the
operators. The global moves are implemented by changing all $\alpha_i$
into a new set of $\alpha^\prime_i$ and accepting the move with a probability
satisfying the detailed balance condition. In this work, we have used
two global moves that switch the spin ($\uparrow \leftrightarrow \downarrow$) and
the orbital ($1 \leftrightarrow 2$) indices.
In the absence of these global moves, the calculation can be trapped
in some regions of the phase space at low temperature, leading to a
wrong (overestimated) value of the polarization.

This is illustrated in Fig.~\ref{fig:dn_vs_beta}, which displays the temperature
dependence of the polarization in the insulating phase, at small $\Delta$. The result of
Eq.~(\ref{eq:polarization_exact}) is compared to the CT-QMC results with and without
global moves. One can see that without global moves, the polarization is bigger than
its strong-coupling value, whereas the contrary is expected. This gives a clear
indication that global moves are needed. When the correct implementation of the
CT-QMC algorithm with global moves is used, the polarization falls below
its strong-coupling value. Note that these results are actually obtained for an intermediate value
of $U=4$, which shows that the range of validity of the strong-coupling approximation is actually quite
extended. 
The agreement between the DMFT data with global moves and the strong-coupling result
is seen to be excellent and both indubitably show the existence of the partially polarized insulator.

\begin{figure}[t]
   \centering
   \includegraphics[clip=true, angle=270, width=.475\textwidth]{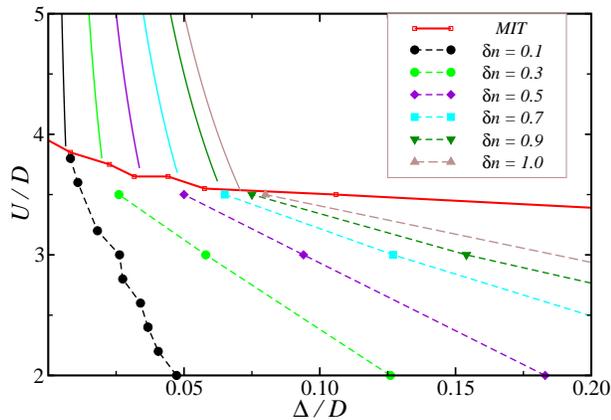}
   \caption{(Color online) Metal-insulator transition (red) and
            iso-polarization lines for the Bethe lattice.
            Different colors mark different values of the polarization (see legend).
            Dashed lines are ED results, while solid lines are the solutions of
            Eq.~\ref{eq:polarization_T=0} for constant polarization.}
   \label{fig:isolines_ext}
\end{figure}

We have not been able (as in Ref.~\onlinecite{PhysRevB.66.115107}),
when using the ED solver at $T=0$ in the insulating phase, to stabilize the
partially polarized insulating solution at small $\Delta$.
This is probably because this solution is too delicate and involves 
a number of competing low-energy
scales ($J$, \D) to be faithfully reproduced given the simple parametrization and
limited number of states in the effective bath, which can be handled within ED in a
two-orbital context.
However, ED performs quite well in the metallic phase, and it is quite
instructive to compare the iso-polarization lines ($\dn=\rm{const.}$) in the
$(\Delta,U)$ plane, determined from ED,
close to the metal-insulator transition, to the
strong-coupling result $\Delta/\Delta_c=\dn$ (i.e., $U \Delta = zt^2 \dn$). This comparison
is made in Fig.~\ref{fig:isolines_ext} in the case of the Bethe lattice (for simplicity).
The ED data on the metallic side of the transition
match very well to the strong-coupling form of the iso-polarization lines
on the insulating side. Thus, this provides a complementary way, starting from the metal, to
document the existence of the PPI regime.

\subsection{Metallic phases and the nature of the metal-insulator transition}
\label{sec:highDelta}

We now turn to the metallic phases. There,
the self-energies can be Taylor expanded at low-frequency as
\begin{equation}
  \label{eq:defZ}
  \Re \Sigma_{\gtrless}(\omega+i0^+) = \Re\Sigma_{\gtrless}(0) + ( 1 - 1/Z_{\gtrless} ) \omega + \cdots,
\end{equation}
in which $Z_>$ and $Z_<$ are the quasiparticle weights of the majority and minority bands, respectively.
The quasiparticle weights, 
$Z_{\gtrless} = (1-\partial\Im\Sigma_{\gtrless}(i\omega)/\partial i\omega)^{-1}|_{\omega \rightarrow 0}$
were extracted from the imaginary frequency data with the use of third-order polynomials.
The minority and majority Fermi surfaces in the metallic phase 
PPM are determined, respectively, (for $V=0$) by:
\begin{subequations}
  \label{eq:eff_mu}
  \begin{eqnarray}
    \label{eq:FS>}
    &&e(\bk)=\mu+\frac{\Delta}{2}-\Re\Sigma_>(0)\equiv\mu_>,  \\
    \label{eq:FS<}
    &&e(\bk)=\mu-\frac{\Delta}{2}-\Re\Sigma_<(0)\equiv\mu_<.
  \end{eqnarray}
\end{subequations}
The quantities $\mu_{>},\mu_<$ can be viewed as effective crystal-field levels
renormalized by interactions (or effective chemical potentials for each type of orbitals),
and a renormalized crystal-field splitting can also be defined as
\begin{equation}
  \label{eq:Delta_eff}
  \Deff\,\equiv\,\Delta+\Re\Sigma_<(0)-\Re\Sigma_>(0).
\end{equation}
The various transitions are conveniently described in terms of $\mu_{\gtrless}$ and
$Z_{\gtrless}$.
On general ground, there are two simple mechanisms by which a given orbital can undergo a transition from a
metallic behavior to an insulating one:

{\it (i)} The quasiparticle weight $Z$ may vanish at the MIT. This is the well-known Brinkman-Rice scenario,
which is realized, e.g., within the half-filled single-band Mott transition within DMFT. It is also
realized for degenerate orbitals with $\Delta=0$: $Z_>=Z_<$ vanishes continuously at
$U_{c}^{\Delta=0}$.

{\it (ii)} It may also happen that either of the equations [Eq.~(\ref{eq:eff_mu})] fails to yield a solution,
i.e., the ``effective chemical potentials'' $\mu_>$ or $\mu_<$ move out of the energy range
$[-D,+D]$ spanned by $e(\bk)$. This, in turn, can happen in a continuous or in a discontinuous
way.

\subsubsection{Orbital polarization and metal-insulator transitions
at large crystal field}
\label{sec:2b1bmetal}

\begin{figure}[!ht]
   \centering
   \includegraphics[clip=true, angle=0, width=.475\textwidth]{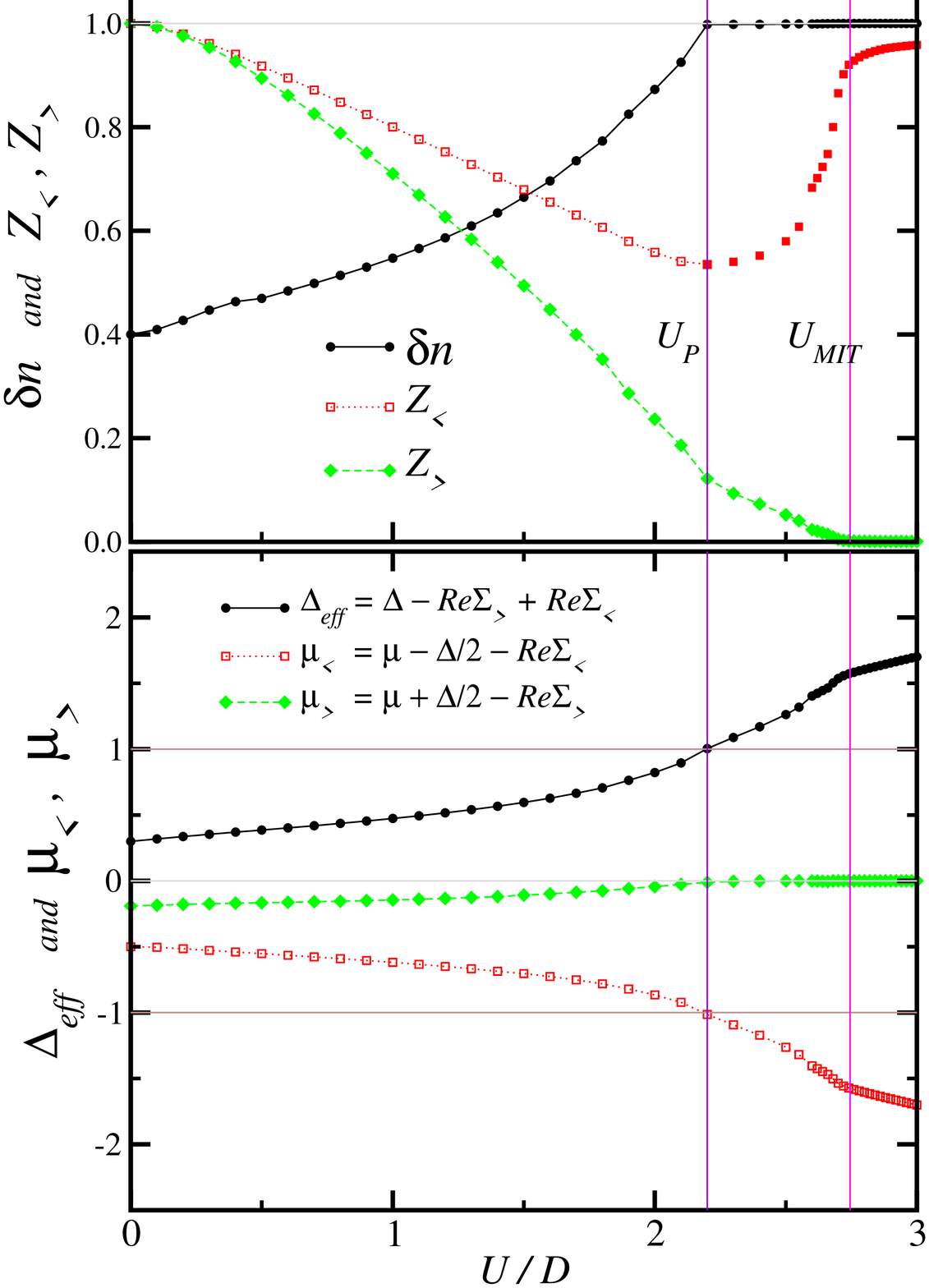}
   \caption{(Color online) PPM-FPM-FPI transitions along the constant $\Delta=0.3$ line
            for the cubic lattice without hybridization ($V=0$).    \\
            Top panel: Orbital polarization, $\delta n$ and QP residues, $Z_>$ and $Z_<$ are shown by
            (black) dots, filled (green) diamonds and open (red) squares, respectively.           \\
             Bottom panel: Effective crystal field splitting, $\Delta_{eff}$ and
             effective chemical potential for both bands, $\mu_>$, $\mu_<$ are represented by
             (black) dots, filled (green) diamonds and open (red) squares, respectively.          \\
             The vertical lines show the full polarization (violet) and MIT transitions (magenta).
             The horizontal (brown) lines show the top and bottom of the bare band. The ED solver was used.  }
   \label{fig:PPM-FPM-FPI}
\end{figure}

We first consider values of the crystal-field splitting larger than $\gtrsim 0.1$.
Two successive transitions are observed as $U$ is increased, from a two-band metal (PPM) to
a single-band metal (FPM) -- followed by a metal-insulator transition (FPM to FPI).
Figure~\ref{fig:PPM-FPM-FPI} (top panel) displays the quasi-particle residues $Z_>,Z_<$
and orbital polarization $\dn$ as $U$ is increased at a fixed $\Delta=0.3$ (indicated by the arrow
on Fig.~\ref{fig:phase_diagram}). The lower panel of Fig.~\ref{fig:PPM-FPM-FPI} displays
$\mu_{\gtrless}$ and $\Deff$.

For $U<U_P (\,\approx 2.2)$, in the two-band metallic phase (PPM),
both quasiparticle weights decrease as $U$ is increased, and the orbital polarization
gradually increases.

At $U=U_P$, the polarization saturates to $\dn=1$ and the minority band becomes
empty. This happens following the mechanism {\it (ii)} above: the minority band effective
level position $\mu_<$ hits the bottom of the band ($\mu_<=-D$ at $U=U_P$) and
the renormalized crystal-field splitting reaches $\Deff=+D$ [as clearly seen
from Fig.~\ref{fig:PPM-FPM-FPI} (lower panel)].  Simultaneously, $\mu_>$ vanishes at $U_P$ and
remains zero for $U>U_P$. This indicates that particle-hole symmetry is restored at
low-energy for the majority band throughout the FPM phase.

\begin{figure}[!ht]
   \centering
   \includegraphics[clip=true, angle=270, width=.475\textwidth]{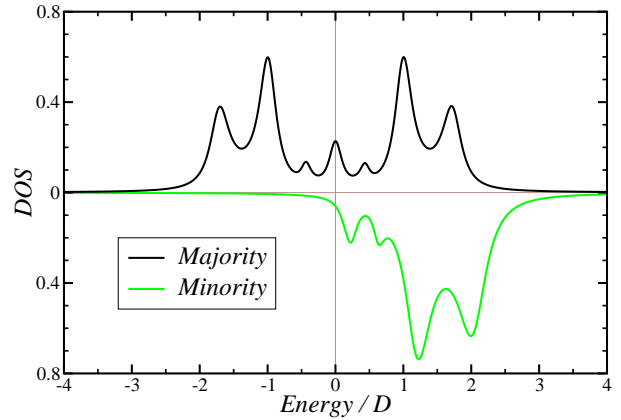}
   \caption{(Color online) Density of states in the 1-band metallic phase.
            The majority (minority) orbital is shown in the upper (lower) half
            of the plot. The ED solver was used with $U=2.4$ and \D=0.3.
            }
   \label{fig:ed_dos}
\end{figure}

For $U_{P}<U<\Umit$, the minority band is empty and becomes inactive. 
The remaining half-filled majority orbital forms a single-band metal and 
is subject to the local Coulomb interaction.
This is illustrated in Fig.~\ref{fig:ed_dos} where we plot the DOS of both orbitals.
Note that the majority orbital very quickly becomes particle-hole symmetric over
its full bandwidth as $U$ increases. The quasi-particle weight of the majority
band $Z_>$ is strongly reduced in this regime.
Note that neither $Z_>$ nor $Z_<$ vanishes at the orbital polarization transition $U_P$.
In fact, also the minority (empty) band self-energy remains linear in frequency at
low energy in this regime, and a $Z_<$ can still be formally defined (as
plotted on Fig.~\ref{fig:PPM-FPM-FPI}), although it no longer has the
physical meaning of a quasiparticle spectral weight since there is no Fermi surface for
that band. In particular, the
increase of $Z_{<}$ in this region should not be interpreted as a
decrease in the correlation effects.

Eventually, the transition from a single-band strongly correlated metal to a
Mott insulator with full orbital polarization is found at $U=\Umit$
($\simeq U_c^{\Delta=\infty}\simeq 2.75$). The nature of this
transition has been exhaustively described in the context of DMFT
studies of the single-band model: $Z_>$ vanishes continuously at the
critical point and the metal-insulator transition is second order (at $T$=0). 
The low-frequency majority self-energy $\Re\Sigma_>(\omega+i0^+)$
acquires a pole on the real frequency axis in the insulating phase. The location
of this pole depends on the choice of the chemical potential within the insulating gap.
For a specific choice (as done in Fig.~\ref{fig:PPM-FPM-FPI}), the pole is located
at zero-frequency so that particle-hole symmetry is restored at low energy and
the self-energy diverges as $\Re\Sigma_>(\omega+i0^+)\sim 1/\omega$.

It should be emphasized that the very small value of $Z_>$ in the
one-band (FPM) metallic phase implies that the quasiparticles are actually quite fragile
in that phase and can be easily destroyed by thermal effects.  Hence,
the orbital polarization transition at $U=U_P$ from a two-band to a one-band metal
at $T=0$ may actually appear, at finite-temperature, as a transition between a
two-band metal and a one-band incoherent ``bad metal'' (or quasi-insulator).
We shall come back to this point in more details in
Sec.~\ref{sec:finiteT}.

\subsubsection{Metal-insulator transition at small crystal field}
\label{sec:MITsmallDelta}

In the small crystal-field regime ($\Delta\lesssim 0.1$), to the best of our numerical
accuracy, there appears to be a simultaneous metal-insulator transition for both orbitals from a two-band
metal (PPM) to a Mott insulator with partial orbital polarization (PPI). Note
that in this region we needed to have recourse to finite-temperature
Monte Carlo simulations.

The nature of the MIT has been well documented by previous DMFT studies
in the degenerate case $\Delta=0$. At $T = 0$, the transition
is second order with a quasiparticle weight $Z_>=Z_<$ vanishing
continuously at $\Uc0$, while at $T > 0$ this transition is first order.

\begin{figure}[!ht]
   \centering
   \includegraphics[clip=true, angle=0, width=.475\textwidth]{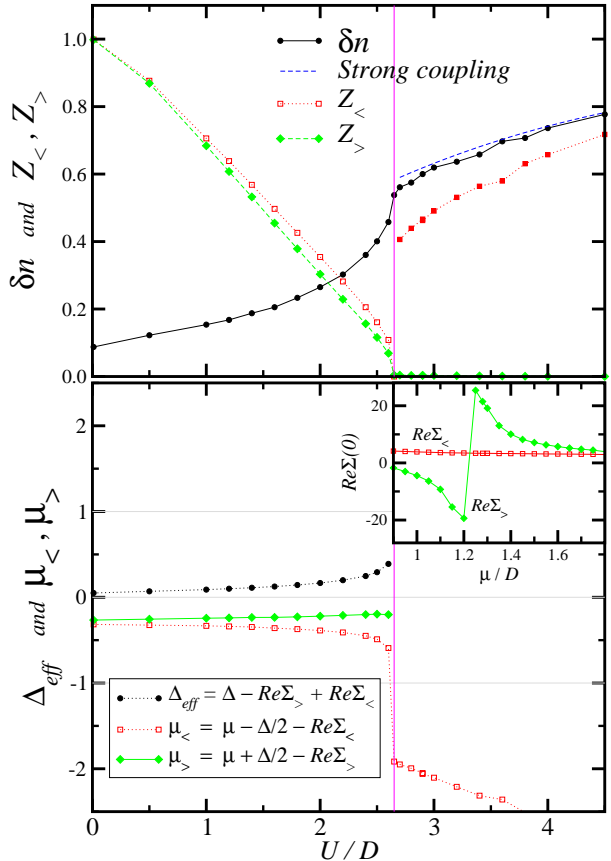}
   \caption{(Color online) PPM-PPI transition along the constant $\Delta$=0.05 line
            for the cubic lattice without hybridization.  \\
            Top panel:
            The (black) solid dots show the orbital polarization, $\delta n$.
            The QP residues, $Z_>$ and $Z_<$, are shown by the (green) filled diamonds and
            (red) open squares, respectively.
            The (blue) dashed line is the strong coupling result.              \\
            Bottom panel: Effective crystal field splitting, $\Delta_{eff}$ and
            effective chemical potential for both bands, $\mu_>$, $\mu_<$ are represented by
            (black) dots, filled (green) diamonds and open (red) squares, respectively.
            The inset shows the real part of the self-energies at the first Matsubara
            frequency, $\Re \Sigma_{\gtrless}(i\pi/\beta)$,
            versus chemical potential within the gap for $U=3$.         \\
            The vertical magenta line shows the MIT. The CT-QMC solver was used with $\beta=100$. }
   \label{fig:PPM-PPI-Transition}
\end{figure}

On the top panel of Fig.~\ref{fig:PPM-PPI-Transition}, we display the quasiparticle weights $Z_>,Z_<$ as a
function of $U$ for $\beta = 100$ and for a small value of $\Delta=0.05$,
along with the orbital polarization $\dn$.
The MIT takes place at a critical coupling $U_c^\Delta$,
which is smaller than $U_{c}^{\Delta=0}$ (Fig.~\ref{fig:phase_diagram}).
Note that the data in Fig.~\ref{fig:PPM-PPI-Transition} is obtained
for a finite temperature and, therefore, the critical $U_{c}^\Delta$
is also smaller than its zero-temperature counterpart (shown in Fig.~\ref{fig:phase_diagram}).
The orbital polarization continuously increases with the interaction 
and does not approach the value $\dn$=1 at the transition point.
The minority orbital quasiparticle weight $Z_<$
remains larger than the majority one $Z_>$ in the metallic phase.
Although it is a delicate issue numerically, our data appear to be consistent with
a majority orbital quasiparticle weight $Z_>$, which vanishes continuously
while $Z_<$ remains finite at the transition.
Note that both the majority and minority orbital effective chemical
potentials [Eq.~(\ref{eq:eff_mu})] stay well within the energy band $[-D,+D]$ for all couplings
in the metallic phase. The transition into the insulating phase for the minority
orbital takes place by having $\mu_<$ jumping out of the energy band in an apparently
discontinuous manner, as we now describe in more details.

After the transition, the chemical potential $\mu$ can be placed (at $T=0$) anywhere within the charge gap,
and therefore, the effective chemical potentials [Eq.~(\ref{eq:eff_mu})] are not longer defined in a unique
manner. As documented in previous work~\cite{Pruschke2005,PhysRevB.55.R4855} 
on the orbitally degenerate case within DMFT, we expect
the majority orbital self-energy to have a pole on the real frequency axis, 
at a position that depends on $\mu$.
For a special choice of $\mu$, this pole is located at $\omega=0$, which should correspond to
a divergence of $\Sigma_>(\omega=0)$ and to a divergent self-energy $\sim 1/\omega$ at low frequency.
In order to document this behavior, we plot in the inset of Fig.~\ref{fig:PPM-PPI-Transition}
the real part of the imaginary frequency self-energies, $\Re \Sigma_{\gtrless}(i\pi/\beta)$
at the first Matsubara point as a function of $\mu$ (for a given value of the interaction $U=3$).
One can clearly see that the majority orbital self-energy,
$\Re \Sigma_>(i\pi/\beta)$ becomes very big and
changes sign at $\mu \sim 1.22$ while $\Re \Sigma_<(i\pi/\beta)$ stays constant within the gap.
A careful scaling analysis shows that $\Sigma_>(\omega = 0)$ indeed diverges at a critical
value of $\mu$. Together with the vanishing of $Z_>$, this shows that the transition for
the majority orbital follows
the standard DMFT scenario identical to the orbitally degenerate case.
Furthermore, since $\Re \Sigma_<(i\pi/\beta)$ does not vary significantly when $\mu$
is varied within the gap, one can unambiguously define $\mu_<$ also in the insulating phase.
In contrast, $\mu_>$ depends on the choice of $\mu$. 
One should note here that 
the chemical potential, $\mu$, defined in this way in the insulating phase continuously
connects to the chemical potential in the metallic phase. 

In Fig.~\ref{fig:PPM-PPI-Transition} (bottom panel), we display these two quantities as a function of $U$,
choosing for $\mu$ the special value at which $\Sigma_>$ behaves as $1/\omega$ at low frequency.
From this plot, we see that the minority band becomes insulating because $\mu_<$ is jumping out of
the energy band in a manner that appears as discontinuous (up to our numerical precision).
Hence, in contrast to the orbital polarization transition of the large $\Delta$ case
described above, the MIT at small $\Delta$ appears to occur in a discontinuous manner,
as far as the minority band is concerned, while being continuous (Brinkman-Rice like) for
the majority band. Note also that the minority-orbital self-energy has a linear behavior at low
frequency throughout the insulating phase.

Note that in this finite-temperature calculation, the orbital polarization
never reaches $\dn = 1$ as $U$ is further increased. 
From the strong-coupling calculation, we expect that it will saturate at $\dn \simeq 0.987$
when $U \rightarrow \infty$. At zero temperature, however, there is a
second-order transition at a finite critical value of $U$ where
the polarization reaches $\dn = 1$.

\section{Effect of an inter-orbital hybridization $V(\bk)$}
\label{sec:finiteV}

\subsection{Low-energy effective-band transition}
\label{sec:effective}

\begin{figure}[t]
   \centering
   \includegraphics[clip=true, angle=270, width=.475\textwidth]{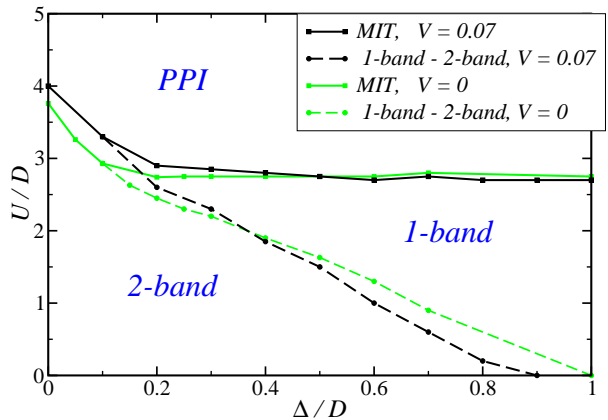}
   \caption{(Color online) Zero-temperature phase diagram 
            of the cubic lattice with hybridization $V=0.07$ and for one electron per site.
            The (black) solid line separates metallic and insulating regions.
            The (black) dot-dashed line separates the 2-band and 1-band metals.
            For the sake of comparison, the corresponding zero-hybridization ($V=0$) lines
            are shown (in green). The ED solver was used. }
  \label{fig:MIT_hyb}
\end{figure}

In this section, we consider the effect of a finite hybridization
(inter-orbital hopping $V(\bk)\neq 0$).
At low values of $\Delta$, the metal-insulator transition is pushed to higher values of
$U$ when turning on a small $V$. While at larger values of $\Delta$, the MIT line is
less sensitive to $V$ (as illustrated on Fig.~\ref{fig:MIT_hyb}).
This is expected since at low $\Delta$
the inter-orbital hopping increases the kinetic energy in both bands while
at higher $\Delta$ the hybridization with a band, which is already empty, has
a smaller effect on the critical coupling.
As we will discuss in more details below, in the presence of
the hybridization, the fully polarized phases (FPM and FPI) disappear.
However, there is still a transition from a two-band to a one-band metal at low
energy. This transition line is pushed up at low values of the crystal-field splitting
because of the increase in kinetic energy.
In non-interacting limit, the finite value of $V$ acts as a $\bk$-dependent
enhancement of the crystal field \D, and therefore, at small values of the
interaction, the two-band to one-band transition line is below the corresponding $V=0$ line.

One should note that the majority (minority) band does not have a unique two
(one) orbital character, and the band index $>$ ($<$) has to be distinguished
from the orbital index two (one).

\begin{figure}[!ht]
   \centering
   \includegraphics[clip=true, angle=0, width=.475\textwidth]{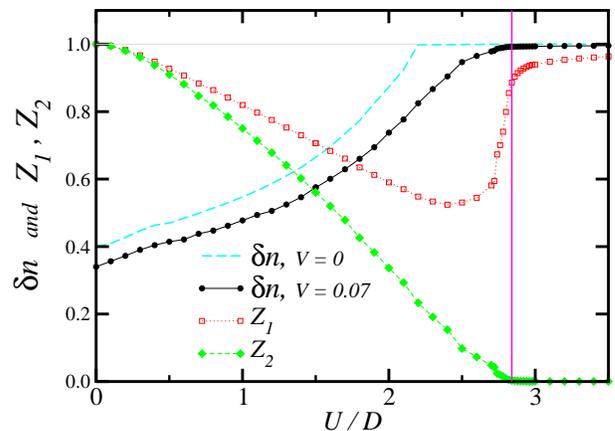}
   \caption{(Color online)
            Orbital polarization $\delta n$ (dots/black), quasiparticle weights $Z_2$
            (green/filled diamonds) and $Z_1$ (red/open squares), as a function of
            $U$ for a fixed value of $\Delta=0.3$ and a finite inter-orbital
            hybridization $V=0.07$.
            For the sake of comparison, the orbital polarization for $V=0$ is also displayed
            (cyan/dashed line).
            The vertical (magenta) line shows the MIT. The ED solver was used.  }
   \label{fig:ED_D=0.3_V=0.07}
\end{figure}

\begin{figure*}[!tbh]
   \centering
   \includegraphics[clip=true, angle=270, width=.95\textwidth]{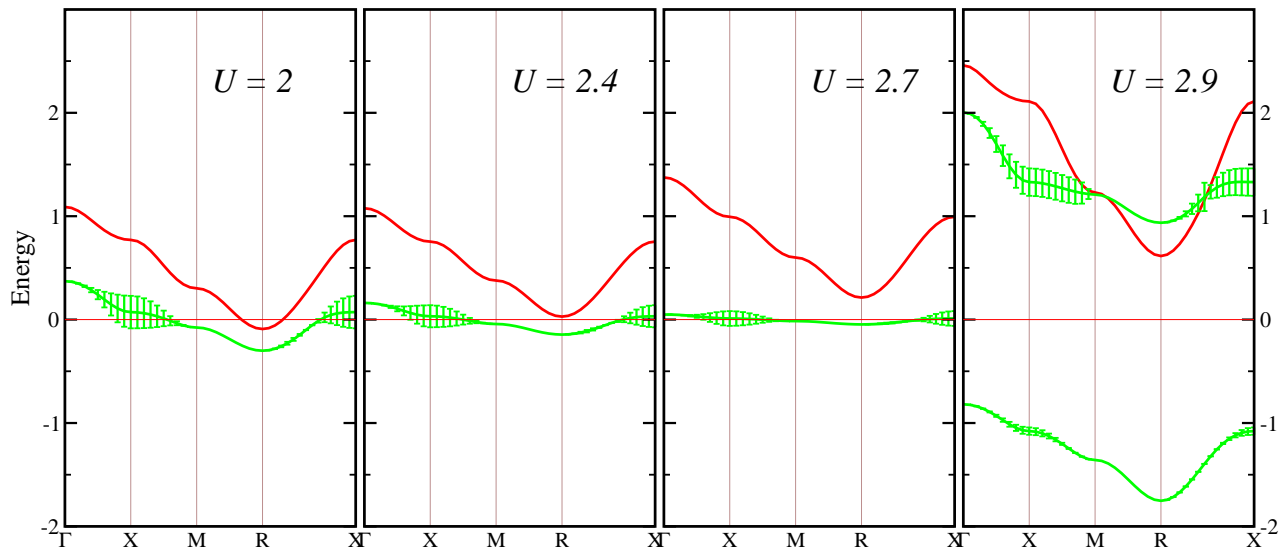}
   \caption{(Color online) Linearized band structure along symmetry lines of the cubic lattice
            for different values of the interaction. We used \D=0.3, $V$=0.07 and the ED solver.
            Fatness shows a contribution of the spectral weight of
            the less occupied orbital ($m=1$) to the majority band (see text for details).
            $U$=2.4 corresponds to the value where the effective crystal field splitting
            exceeds the bare bandwidth and the physical picture effectively becomes
            single band. The rightmost panel corresponds to the PPI solution and
            we used $\Re\Sigma_2(\omega+i0^+)=\Re\Sigma_2(0)+\Omega/\omega$ for
            the divergent orbital.}
   \label{fig:ED_bands}
\end{figure*}

On Fig.~\ref{fig:ED_D=0.3_V=0.07}, we display the quasiparticle weights and orbital polarization
as a function of $U$, for a fixed value of $V$ and a rather large crystal field
$\Delta=0.3$. One clearly sees that the MIT follows a similar mechanism than in
the $V=0$ case: only $Z_2$ vanishes continuously at the transition, while $Z_1$ is always finite.

A noticeable difference with the $V=0$ case is that the orbital polarization $\dn=n_2-n_1$
does not reach saturation ($\dn<1$) before the MIT (Fig.~\ref{fig:ED_D=0.3_V=0.07}). This is
expected, because the low-energy bands in the metallic state no longer have a unique ($1,2$) orbital
character, as we now discuss.

In order to understand more precisely the nature of the metallic phase,
we use the low-frequency expansion of the self-energies and we obtain the expressions
of the low-energy majority and minority bands, which read,
\begin{subequations}
  \begin{eqnarray}\label{eq:min}
    &&2\omega_{<}(\bk)=\\ \nonumber
    &&= Z_1\e1+Z_2\eps2 + \sqrt{(Z_1\e1-Z_2\eps2)^2+4Z_1Z_2\vk^2}, \\ \label{eq:maj}
    &&2\omega_{>}(\bk)=\\ \nonumber
    &&= Z_1\e1+Z_2\eps2 - \sqrt{(Z_1\e1-Z_2\eps2)^2+4Z_1Z_2\vk^2}.
  \end{eqnarray}
\end{subequations}
In these expressions $\e1\equiv \ek-\mu+\Delta/2+\Re\Sigma_1(0)$ and
$\eps2\equiv \ek-\mu-\Delta/2+\Re\Sigma_2(0)$. The Fermi surface
(set by $\omega=0$) is determined by the following condition (in which
the weights $Z_{1,2}$ do not appear):
\begin{eqnarray}
  \label{eq:FS}
   0 & =      & \e1 \eps2 - \vk^2                                      \\          \nonumber
     & \equiv & \left[\ek-\mu+\Delta/2+\Re\Sigma_1(0)\right] \times     \\          \nonumber
     &        & \left[\ek-\mu-\Delta/2+\Re\Sigma_2(0)\right]  - \vk^2.
\end{eqnarray}
We recall that, when $V=0$, an orbital polarization transition is first encountered at
$U=U_P$, at which the Fermi-surface sheet corresponding to orbital $1$ (determined by
$\e1=0$) disappears, since $\mu-\Delta/2-\Re\Sigma_1(0)$ reaches the band-edge. In the presence
of $V\neq 0$, a similar phenomenon occurs for the minority low-energy band
$\omega_<(\bk)$: one of the two sheets, which constitute the solution of Eq.~(\ref{eq:FS}) ceases
to exist. This is expected from continuity arguments in view of Eq.~(\ref{eq:FS}) and of
the situation at $V=0$. This is furthermore demonstrated by Fig.~\ref{fig:ED_bands},
which displays the majority and minority low-energy bands $\omega_{\gtrless}(\bk)$
along the main directions in the Brillouin zone, as
$U$ is increased. It is clearly seen from this figure that for $U\simeq 2.4$ (before
the MIT, which takes place at $U\simeq 2.8$), an effective band transition occurs between
the two-band metal and a one-band metal at low energy. The critical coupling for this
effective-band transition is slightly increased as compared to its
value at $V=0$.

For $V=0$, the majority eigenstate $|\bk,>\rangle$ (corresponding to eigenvalue
$\omega_>(\bk)$) has a unique orbital character $m=2$. In contrast, for $V\neq 0$, it
has a component on both orbital $2$ and orbital $1$. As a result, the orbital polarization does
not reach $\dn=1$ (Fig.~\ref{fig:ED_D=0.3_V=0.07}) at the effective band transition between a
two-band and a one-band metal. On Fig.~\ref{fig:ED_bands}, we have used a ``fat band'' representation to
illustrate this point: at each $\bk$ point, we plot a bar whose extension is proportional to the
matrix element $|\langle 1 | \bk,> \rangle|^2$, measuring the projection of the less-occupied
orbital $m=1$ onto the majority band.

As $U$ is increased beyond the effective-band transition, one is left with
a single effective low-energy band, characterized by a quasiparticle weight,
\begin{equation}
  Z_> (\bk) = \frac{(\e1 + \eps2) Z_1 Z_2}{\e1 Z_1 + \eps2 Z_2},
\end{equation}
where $\bk$ lies on the Fermi surface of the majority band [se Eq.~(\ref{eq:FS})].
The subsequent Mott transition is characterized by a
vanishing quasiparticle weight for the majority band $Z_> \sim Z_2 \rightarrow 0$, as
clearly seen from Fig.~\ref{fig:ED_D=0.3_V=0.07} and from the narrowing of that
band in the third panel of Fig.~\ref{fig:ED_bands}.

The key conclusion of this section is that, even in the presence of a finite inter-orbital
hopping, two distinct transitions are observed as $U$ is increased (in the large crystal-field
regime): first, a second-order transition from a metal with two active bands at low energy to
a metal with only one active band at low energy, and followed by a Mott metal-insulator
transition of the one-band type.

\subsection{Orbital-selective coherence and the two-band metal to one-band bad-metal transition}
\label{sec:finiteT}

\begin{figure}[!ht]
   \centering
   \includegraphics[clip=true, angle=0, width=.475\textwidth]{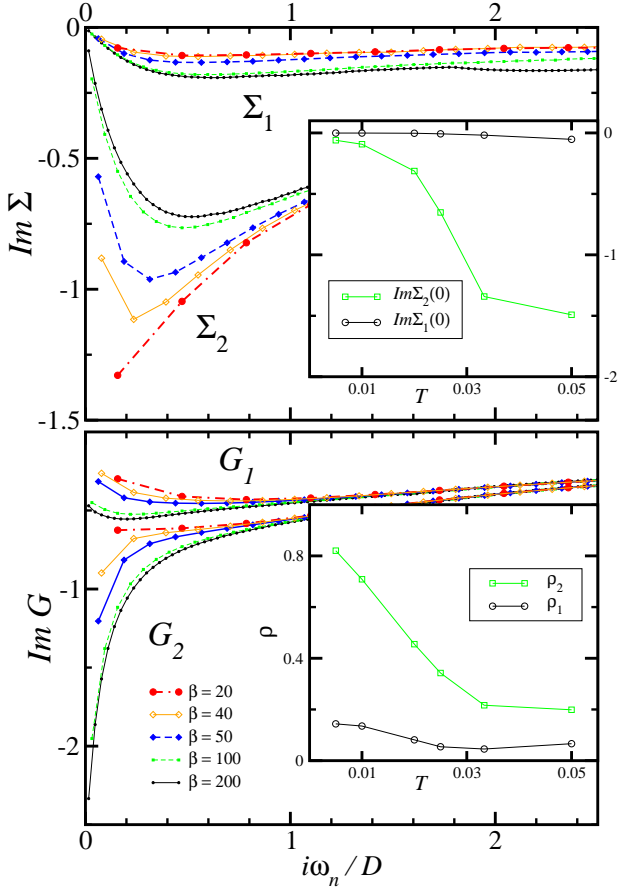}
   \caption{(Color online) Top panel: Imaginary part of the self-energies,
            $\Im \Sigma_1(i\omega_n)$ and $\Im \Sigma_2(i\omega_n)$ for different temperatures
            (see legend for temperature coding).
            The inset shows the extrapolation to zero of the imaginary part of the self-energies,
            $\Im \Sigma_{1,2}(0)$ versus temperature.                   \\
            Bottom panel: Imaginary part of the Green's functions,
            $\Im G_1(i\omega_n)$ and $\Im G_2(i\omega_n)$ for different temperatures
            (the color-coding is the same).
            The inset shows the density of states at the chemical potential
            $\rho_{1,2}(0)$ versus temperature.
            We used $U=2$, $\Delta=0.3$, $V=0.07$ and the CT-QMC solver.}
   \label{fig:sigma_U=2}
\end{figure}

We have seen above that, in a rather extended region of the metallic phase,
the quasiparticle weight of the majority orbital is much smaller than that of the
minority one. This is especially true close to the two-band to one-band metal transition,
where $Z_2 \ll Z_1$. This implies that thermal effects can easily destroy the fragile
quasiparticles of the majority band.
This has physical consequences, which may be important in practice. For example, the two-band metal
to one-band metal transition at finite temperature may appear in practice as 
a quasi-metal-insulator transition or more precisely
as a transition between a two-band metal and a bad (or incoherent) metal. This will happen
when the temperature, at which the system is studied, is higher than the (small)
quasiparticle coherence temperature of the majority band.

In order to illustrate this point, we performed finite-temperature studies for the following
parameter values: $U=2$, $\Delta=0.3$, and $V=0.07$, which correspond to the two-band metallic regime,
not very far from the two-band to one-band metal transition. For these parameters, the
two quasi particle residues are $Z_2$=0.34 and $Z_1$=0.59 (see Fig.~\ref{fig:ED_D=0.3_V=0.07}).
In Fig.~\ref{fig:sigma_U=2}, we display the imaginary part
of the Green's functions (bottom) $\Im G_{1,2}(i\omega_n)$ and self-energies (top)
$\Im \Sigma_{1,2}(i\omega_n)$
on the Matsubara axis, for different temperatures. In the insets of this figure, we display the
extrapolated zero-frequency value $\Im \Sigma_{1,2}(i0^+)$, which is related to
inverse quasiparticle lifetime and zero-frequency density of states,
$\rho_{1,2}(0) \equiv - \Im G_{1,2}(i0^+)/\pi$, respectively.

It is seen from these figures that, while the minority orbital quantities have quite
little temperature dependence, the majority orbital, in contrast, displays very strong temperature
dependence. For example for $T\gtrsim 0.03$ (i.e., a rather low-energy scale as compared to the
bandwidth), the majority orbital is clearly incoherent with a small quasiparticle lifetime
and much reduced value of the local density of states. At those temperatures, the frequency dependence of
the self-energy is clearly non-metallic, extrapolating to a large value at zero frequency. Only
at a low temperature $T\sim 0.01$ ($200$ times smaller than the bandwidth), the behavior of
a coherent metal is recovered, with a linear Matsubara frequency dependence of $\Im\Sigma_2(i\omega_n)$
extrapolating to a small value at low frequency (corresponding to a large quasiparticle
lifetime).

\section{Conclusion}
\label{sec:conclusion}

In this paper, we have investigated how a crystal-field splitting, by
lifting orbital degeneracy, affects the Mott metal-insulator transition in
the presence of strong on-site correlations. The study was performed on a
simple two-orbital model at quarter filling (one electron per site), and we
have also considered the effect of an inter-orbital hopping (hybridization),
which is important for applications to real materials. 

Within the metallic
phase, a second-order transition from a two-band to a one-band metal takes place as the
crystal field is increased. 
The critical value of the crystal-field
splitting, at which this transition takes place, is considerably lowered for
strong on-site repulsion (i.e., the effective crystal-field splitting is
considerably enhanced). 
This transition has the nature of a effective band
transition for the renormalized low-energy bands (i.e., the minority band is pushed up
in energy and does not cross the Fermi energy anymore)
and survives in the presence of an inter-orbital hopping. 

The nature of the Mott
metal-insulator transitions induced by on-site repulsion was found to
depend on the magnitude of the crystal-field splitting. At high enough
values of this splitting, the Mott transition is between a one-band metal
and a one-band Mott insulator (conventional Brinkman-Rice scenario): 
only the majority orbital is involved, and
the transition is second order and characterized by a vanishing
quasi-particle weight for that orbital. At low values of the crystal-field
splitting, the transition is from a two-band metal to a Mott insulator with
{\it partial orbital polarization}. It takes place simultaneously for both
orbitals: although the transition is still continuous for the majority
orbital, it has a first-order character for the minority orbital.
Elucidating these transitions and, in particular, establishing the
existence of the partially orbitally polarized Mott insulator at low
crystal fields was made possible by the recent development of the CT-QMC
algorithm for the solution of the DMFT equations.

If a finite hybridization ($V\neq0$) is taken into account, it is no longer
possible to fully polarize the system. Therefore, the FPM and the FPI phases
disappear. However, there is still a transition from a two-band to a one-band metal
at low energy so that the introduction of a finite $V$ does not modify the
overall picture of the model.

We have also studied the influence of the temperature on the two-band metal
just below the transition to the one-band metal. The temperature can
easily drive the system into a regime where the quasiparticle
weight of the majority band is destroyed and the system effectively becomes 
a single-band metal. Further increase of the temperature above the
characteristic temperature of both bands leads the system
into an incoherent (or bad) metal.

Our study has direct relevance for the interpretation of the metal-insulator transitions of
transition-metal oxides (see Sec.~\ref{sec:intro}), often accompanied by an enhanced orbital
polarization.

\acknowledgements

We are very grateful to O.K.~Andersen, S.~Biermann, A.~Rubtsov, and 
A.~Lichtenstein for the discussions related to this work. 
We also thank V.~Anisimov, F.~Lechermann, A.~Millis, and P.~Werner for the useful conversations.
We acknowledge the support from CNRS, Ecole Polytechnique, the Agence Nationale de la
Recherche (under contract {\it ETSF}), and the Marie Curie Grant No. MIF1-CT-2006-021820.
This work was supported by a supercomputing grant at IDRIS Orsay under Project No. 071393
(for the ED results) and at CEA-CCRT under Project No. p588 (for the CT-QMC calculations).

\bibliography{xfield}

\end{document}